# An Extended Gate Field Effect Transistor (EGFET)-based pH Microsensor Utilizing a Polypyyrole Coated pH Microelectrode

Mustafa Şen[a,××], Fikri Seven[b]

[a]Department of Biomedical Engineering, Izmir Katip Celebi University, Izmir, Turkey.
[b]Department of Biomedical Engineering, Izmir Katip Celebi University, Izmir, Turkey.

ABSTRACT

Here, an ultra-small size, simple and inexpensive metal oxide semiconductor field effect transistor (MOSFET)-integrated needle type EGFET pH microsensor was fabricated. The EGFET pH microsensor has the potential to be applied to fast and precise local pH measurements. The system was composed of two components; a pH sensitive probe and a MOSFET. The pH sensitive probe was made by electrochemically coating the surface of a Pt ultra micro-electrode with polypyrrole, a semiconductor polymer. The pH sensitive probe was then integrated with the gate of the MOSFET to carry out measurements in PBS at different pH values. The real time response of the EGFET pH microsensor was also tested by cycling the probe in three solutions at different pH. The results showed that the developed pH microsensor is sensitive to pH change. It is expected that the EGFET pH microsensor will allow local pH analysis in biological samples or corrosion studies.

## 1. Introduction

Microelectrode probes are used in local measurements and electrochemical imaging in many different areas such as corrosion [1], single cell studies [2] and measurement of neurotransmitter release [3]. There are various studies in the literature on the fabrication of such type of microprobes that allow local pH measurement. The fact that traditional measurement methods such as glass pH electrodes could not be miniaturized for the application at the desired level in micro scale led to the production of these micro probes with new approaches. Generally, pH microprobes have disadvantages such as limited pH measurement range, long response time for meaningful signal, and sensitivity. H+ ion selective field effect transistors (FET) [4], optical pH sensors [5] and microelectrode pH sensors [6] are more outstanding examples and have been used successfully in certain studies in the literature. Although ion-selective FET (ISFET) based pH sensors are relatively easy to miniaturize, they have disadvantages such as low sensitivity and long signal stabilization time. Optical pH sensors do not need a reference electrode and can successfully detect the spatial distribution of H+ ions in sea water [7]. However, since such sensors can be severely affected by the color of the solution, they are used in almost colorless solutions whose transparency does not change significantly for an accurate result. Microelectrodes are considered powerful tools for in vivo or on-site measurements in various fields such as physiology, microbial ecology, medicine, neuroscience, and environmental monitoring. In the literature, there are various studies on the effective use of glass pH microelectrode pH sensors in local pH measurement due to their advantages such as high selectivity, reliability and wide dynamic range. These types of pH sensors are mostly used as scanning electrochemical microscopy (SECM) probe and due to the structure of this microscope, these probes can display local pH distributions.

In this study, an ultra-small size, simple and inexpensive EGFET pH microsensor has been fabricated with a new approach. In short, the EGFET pH microsensor was developed by integrating a metal oxide semiconductor FET (Metal Oxide Semiconductor Field Effect Transistor - MOSFET) and a Pt ultramicro electrode (UME) coated with a semiconductor polymer, polypyrrole. The EGFET pH microsensor was designed for fast and precise local pH measurement. It is anticipated that the EGFET pH microsensor can be used for local pH measurement in small-volume environments, biological samples or corrosion studies.

## 2. MATERIAL AND METHOD

### 2.1. Material

Phosphate buffered saline (PBS) (Sigma Aldrich, USA), NaOH (Sigma-Aldrich, USA), HCl (Sigma-Aldrich, USA), pyrrole (Sigma-Aldrich, USA), Ag paste (Sigma-Aldrich, USA), n-type MOSFET (IRFZ44N, International Rectifier, USA), Pt wire (Sigma-Aldrich, USA), glass capillary tube (World Precision Instruments, USA).

### 2.2. Fabrication of Pt ultramicroelectrodes (Pt-UME)

Platinum microelectrodes can be fabricated in various ways and in this study, it was fabricated by inserting a Pt wire into the glass capillary tube [8, 9, 10, 11, 12, 13]. Pt wire, with a diameter of 25 $\mu$m, was connected to a copper wire covered with an open non-conductive material before this process, so that the Pt wire could be easily placed inside the single glass capillary tube. For the bonding process, the Pt and Cu wires were bonded to each other by using silver (Ag) paste and then the connection was made permanent by the solidification of the silver paste at 180 ° C. Next, the Pt wire was placed inside the single glass capillary tube with the help of the copper wire to which it was fixed, and the glass capillary tube was pulled over the Pt wire with a micropuller (PC-10, Narishige, Japan) machine. With the help of a "microforge" (MF-830, Narishige, Japan) machine, the gap between the glass and the Pt wire was removed and the probe was produced in a leak-free manner. Finally, by grinding the tip of the probe using a micro grinder (Microgrinder, Narishige, Japan), the excess Pt wire in the probe tip was removed from the environment and the electrode was produced as a micro-disc.

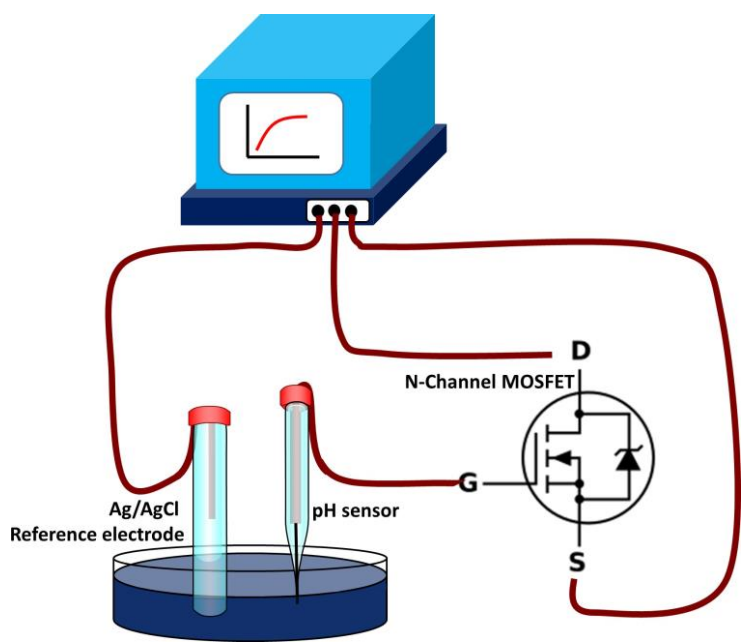

**Figure 1:** EGFET pH microsensor constructed through the integration of a polypyrrole coated Pt microelectrode and a MOSFET

### 2.3. Deposition of pH sensitive polymer on Pt-UME

The surface of the Pt-UMEs produced in this section is covered with a semiconductor polypyrrole polymer by electrochemical deposition method. Cyclic voltammetry was used for coating the polymer. Briefly, the potential applied to Pt-UME immersed in 0.1 mM pyrrole solution was scanned 4 times between 0 and +0.8 V (vs. Ag / AgCl). [14].

### 2.4. Use and characterization of EGFET pH microsensor in pH measurement after integration with MOSFET

The pH measurement capacity of the microprobes using FET analysis unit (B2901A Precision Source / Measure Unit) was demonstrated and characterized in solutions with different pH values. The source and drain of the n-type MOSFET field effect transistor used for measurement are connected to the source and drain inputs of the FET analysis unit, while the gate part is connected to the produced EGFET pH microsensor (Figure 1). The pH measurement was carried out in the presence of Ag / AgCl reference electrode to make the measurement more precise and stable. The pH sensor and Ag / AgCl reference electrode components of the prepared system were immersed in solutions with different pH values and measurements were carried out. I-V curves were obtained for measurement. Briefly, while applying a constant potential (+1 V) to the gate, the potential applied between the source and the drain was scanned in a wide range (from 0 to +1 V). The sensitivity of the pH sensor fabrication after the characterization process was determined and its measurement capacity was displayed in different solutions. The basic starting point here is that the polymer, which is sensitive to pH change, causes changes in the curves obtained by manipulating the effect of the potential applied to the gate on the channel in solutions with different pH values. Measurements were carried out in PBS solutions with pH values of 2, 4, 6, 8 and 10. The time-dependent change is very important in terms of displaying the potential of the produced pH sensor, especially in pH monitoring as SECM probe. At this point, the EGFET pH microsensor was cycled between three solutions at different pHs (4.68, 6.8 and 9.5).

## 3. Results

Pt-UMEs have been successfully fabricated using the proposed technique. Pt-UMEs were characterized electrochemically before use in the micro-pH sensor. The CV curves obtained from 0 to + 0.5 V (Ag / AgCl) in 1 mM FcCH2OH solution were found to be compatible with the theoretically obtained data, and therefore it was concluded that these probes produced were leak-proof and could be used in numerical analysis. The CV curve obtained during polypyrrole coating was also as expected, and the current value obtained at +0.8 V after each coating showed a decreasing trend. The produced Pt-UME was connected to an n-type MOSFET as shown in Figure 1 and I-V curves were taken in PBS solutions with different pH values. As seen in Figure 2A, the obtained I-V curves differed depending on the pH of the solution being measured. The change in the conductivity of the semiconductor polypyrrole at different pH has modulated the potential applied to the channel between the source and the drain in the MOSFET. This situation resulted in obtaining different I-V curves at different pH values. The calibration curve based on $I_D$ values obtained in different pH solutions at +1 $V_{DS}$ shows that the relationship between pH and $I_D$ is linear ($R^2 = 0.9963$). The sensor had a pH sensitivity of 0.4 mA/pH. When the EGFET pH microsensor was cycled in solutions at different pHs (pH 4.68, 6.8 and 9.5), the $I_D$ response gained a steady state in a matter of seconds. The hyteresis was calculated to be 0.06 mA. These result prove that the response of EGFET pH microsensor in relatively wide range of pH 2 to 10 is accurate. Microelectrodes have several advantages over traditional macroscale electrodes, such as fast response, high current density, high signal-to-noise ratio, low i-R drop, small double layer capacitance. Microelectrodes are considered powerful tools for in vivo or on-site measurements in various fields such as physiology, microbial ecology, medicine, neuroscience, and environmental monitoring. In the literature, there are various studies on the effective use of glass pH microelectrode pH sensors in local pH measurement due to their advantages such as high selectivity, reliability and wide dynamic range [15]. This type of pH sensors are mostly used as scanning electrochemical microscopy (SECM) probe and can display local pH distributions with these probes due to the structure of this microscope. izquierdo2013potentiometric. Electrochemical signals in SECM experiments are very sensitive to tip-substrate separation [16] and pH measurement in smaller scales requires a very precise control of the probe position. The biggest handicap of microelectrode pH sensors is their short lifetime and for that they are mostly used in the analysis of biological samples [17]. It is thought that the present EGFET pH microsensors have a high potential of use in areas where local analysis is important, especially in tissue engineering and single cell studies.

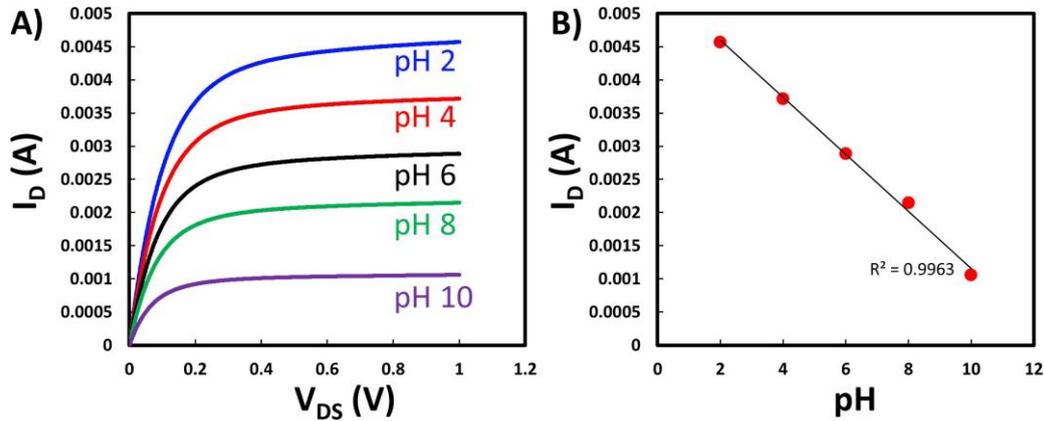

**Figure 2:** I-V curves obtained at different pH values (A). Calibration curve (B) obtained using $I_D$ values measured at +1 $V_{DS}$ for each pH level.

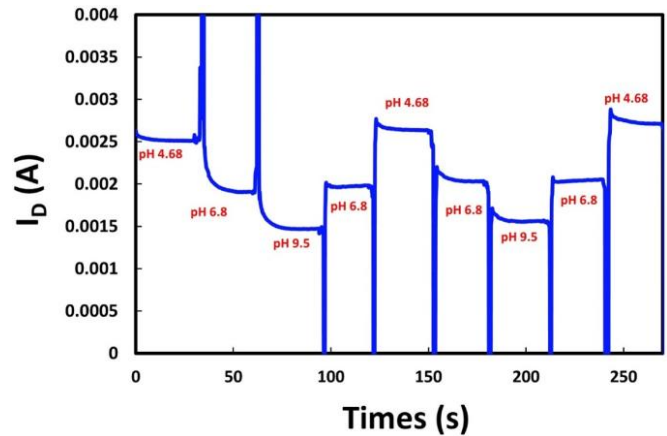

**Figure 3:** The real-time $I_D$ response of the EGFET pH microsensor at +1 $V_{DS}$ and $V_{GS}$ when cycled between in PBS solutions at pH 4.68, 6.8 and 9.5.

## Acknowledgement

This study was carried out as a part of the project numbered 2017-ÖDL-MÜMF-0010, supported by İzmir Katip Çelebi University Scientific Research Projects.